\documentclass[a4paper,11pt]{article}
\usepackage{pos}

\usepackage[font=small, labelfont=bf, textfont={small,it}]{caption}
\usepackage{slashed}
\usepackage{braket}
\usepackage[utf8]{inputenc}
\usepackage[greek,english]{babel}
\usepackage{wrapfig}
\graphicspath{{figs/}}

\DeclareFontFamilySubstitution{LGR}{ntxtlf}{Tempora-TLF}

\newcommand{\beq}{\begin{eqnarray}}
\newcommand{\eeq}{\end{eqnarray}}
\newcommand{\beqnn}{\begin{eqnarray*}}
\newcommand{\eeqnn}{\end{eqnarray*}}
\newcommand{\pp}{{\scriptscriptstyle{(+)}}}
\newcommand{\mm}{{\scriptscriptstyle{(-)}}}
\renewcommand{\S}{{\scriptscriptstyle{\mathrm{S}}}}
\newcommand{\five}{{\scriptscriptstyle{5}}}
\newcommand{\QCD}{{\scriptscriptstyle{\mathrm{QCD}}}}
\newcommand{\clov}{{\scriptscriptstyle{\mathrm{clov}}}}
\newcommand{\cool}{{\scriptscriptstyle{\mathrm{cool}}}}
\newcommand{\f}{{\scriptscriptstyle{\mathrm{f}}}}

\newcommand{\E}{{\scriptscriptstyle{\mathrm{E}}}}
\renewcommand{\L}{{\scriptscriptstyle{\mathrm{L}}}}
\newcommand{\Left}{{\scriptscriptstyle{\mathrm{L}}}}
\newcommand{\Right}{{\scriptscriptstyle{\mathrm{R}}}}
\newcommand{\light}{{\scriptscriptstyle{(\mathrm{\ell})}}}
\newcommand{\phys}{{\scriptscriptstyle{(\mathrm{phys})}}}
\newcommand{\dd}{\mathrm{d}}
\newcommand{\ee}{\mathrm{e}}
\newcommand{\ii}{\mathrm{i}}
\newcommand{\Tr}{\mathrm{Tr}}
\newcommand{\rhobar}{\overline{\rho}}
\newcommand{\omegabar}{\overline{\omega}}
\newcommand{\Rs}{R_{\scriptscriptstyle{\mathrm{s}}}}

\title{From strong interactions to Dark Matter: the non-perturbative QCD sphaleron rate}
\ShortTitle{From strong interactions to Dark Matter: the non-perturbative QCD sphaleron rate}

\author*[a,b]{Claudio Bonanno}

\affiliation[a]{Instituto de F\'isica Te\'orica UAM-CSIC, Calle Nicol\'as Cabrera 13-15,\\Universidad Aut\'onoma de Madrid, Cantoblanco, E-28049 Madrid, Spain}

\affiliation[b]{Albert Einstein Center for Fundamental Physics, Institute for Theoretical Physics, University of Bern, Sidlerstra\ss e 5, CH-3012 Bern, Switzerland}

\emailAdd{claudio.bonanno@unibe.ch}

\abstract{Acceptance plenary talk for the 2025 Kenneth G.~Wilson Award for Excellence in Lattice Field Theory: \emph{For significant contributions to the understanding of topology in QCD, QCD-like, and large-$N_c$ gauge theories, including algorithmic developments to reduce topological freezing, studies of Dirac spectral properties, and axion phenomenology.}}

\FullConference{The 42$^{\text{nd}}$ International Symposium on Lattice Field Theory (LATTICE2025)\\
2--8 November 2025\\
Tata Institute of Fundamental Research, Mumbai, India\\}

\begin{document}
\maketitle

\section*{Kenneth G.~Wilson Award}

Having been selected as the recipient of the 2025 Kenneth G.~Wilson Award (KWA) for Excellence in Lattice Field Theory is an immense honor. When I received the news, I could not help but thinking about how exciting it is to be part of the large and diverse community working on Lattice Field Theory, initiated by Wilson's fundamental work on confinement. What I find particularly stimulating is that, being Lattice Field Theory sitting at the crossroad between Physics, Mathematics and Computational Sciences, many different expertise had to come together to make progress in this field possible. In my professional experience so far, I have done my best to give my own contributions to these three facets of Lattice Field Theory --- theory, phenomenology, and algorithms. I am extremely grateful to the KWA Selection Committee, to the International Advisory Committee, and to the Lattice 2025 conference organizers, for acknowledging my contributions. I am also extremely grateful for all the warm and sincere expressions of esteem I have received after this recognition from many other colleagues.

Clearly, I would have never reached this achievement without the invaluable role of my collaborators, and without the excellent work environment provided by Pisa, Arcetri, Firenze, Madrid, and now Bern. I have been blessed with the possibility of being part of fruitful scientific environments that fostered my professional development, and of collaborating with many excellent scientists. I would like to extend my heartfelt thanks to all my collaborators and to all the colleagues I have scientifically interacted with for making research such a stimulating and positive experience. I will not list all of you, but I must make two exceptions. First, I want to thank Massimo D'Elia for having been the best professor, supervisor, mentor, and collaborator, over the almost-decade we have been working together. Your sincere encouragement and support helped me to find my own way in research. In the years to come, I aspire to follow the model of mentorship and supervision you represent. Then, I want to thank Margarita Garc\'ia P\'erez for making my three years in Madrid one of the most formative experiences of my life. The countless discussions about physics, and beyond, with you made me grow both as a person and as a researcher. You have taught me so many things --- including Spanish, by talking to me in your beautiful language --- and I would like to express my gratitude to you for being such a wonderful person and collaborator. !`Muchas gracias!

\vspace{0.5\baselineskip}

In what follows, I will propose a transcript of the KWA acceptance plenary talk I have given at Lattice 2025. In that contribution, I decided to present the main findings of the papers~\cite{Bonanno:2023ljc,Bonanno:2023thi,Bonanno:2024zyn} for three main reasons. First, they constitute my and my collaborators' latest contributions to the topics mentioned in the award. Secondly, I believe they can be of broad interest both to the Lattice community and to scientists working in neighboring fields. Finally, they allow me to delineate what are possible future directions to further develop the field of investigation cited in the award.

\section{Theoretical and phenomenological implications of the strong sphaleron rate}

Since the pioneering works of Witten~\cite{Witten:1979vv} and Veneziano~\cite{Veneziano:1979ec}, it has been clear that gauge topology plays a prominent role in determining many of the salient theoretical features of gauge theories, with fundamental implications for phenomenology. Topological properties of gauge theories stem from purely non-perturbative dynamics, and Lattice Quantum Chromo-Dynamics (QCD) has been instrumental to address this topic from first principles. By now, there are several topics in gauge topology that have been extensively addressed from the lattice, such as $\theta$-dependence and its connection with the Witten--Veneziano mechanism~\cite{DelDebbio:2002xa,DelDebbio:2004ns,Bonati:2015sqt,Ce:2015qha,Bonati:2016tvi,Ce:2016awn,Athenodorou:2020ani,Athenodorou:2021qvs,Bonanno:2025eeb} and with axion phenomenology~\cite{Bonati:2015vqz,Petreczky:2016vrs,Borsanyi:2016ksw,Trunin:2015yda,Burger:2017xkz,Bonati:2018blm,Burger:2018fvb,Lombardo:2020bvn,Athenodorou:2022aay,Kotov:2025ilm}, the neutron electric dipole moment~\cite{Shindler:2015aqa,Guo:2015tla,Dragos:2019oxn,Alexandrou:2020mds,Bhattacharya:2021lol,Liang:2023jfj,Bhattacharya:2023qwf}, the role of topology in the QCD phase diagram~\cite{DElia:2012pvq,DElia:2013uaf,Otake:2022bcq,Bonanno:2023hhp,Yamada:2024pjy}, and more. This article is about a topic in gauge topology that has received far less attention from the lattice: \emph{real-time sphaleron transitions} in strong interactions.

\begin{wrapfigure}{r}{0.5\textwidth}
\centering
\includegraphics[scale=0.25]{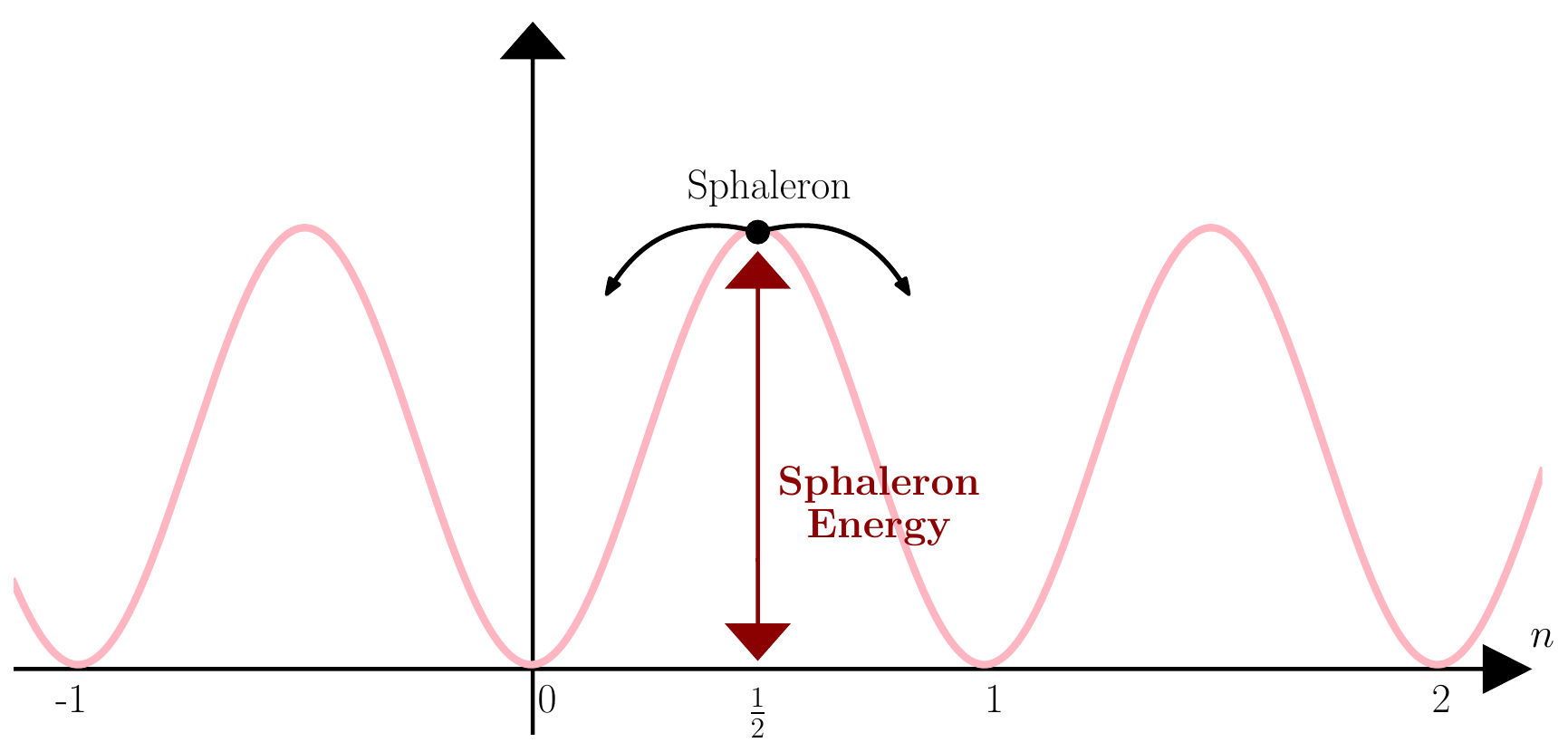}
\caption{Semiclassical cartoon of the landscape of topologically-non-equivalent Yang--Mills vacua, labeled by an integer winding number $n$, and of a sphaleron transition above the potential barrier.}
\label{fig:sphaleron_cartoon}
\end{wrapfigure}

Sphalerons are unstable field configurations corresponding to saddle-points of the Yang--Mills action in Minkowski time~\cite{Christ:1979zm,Klinkhamer:1984di,Klinkhamer:2017fqi}. At the semiclassical level, they can be interpreted as describing transitions among topologically-non-equivalent vacua due to thermal fluctuations above potential barriers separating them, see the cartoon in Fig.~\ref{fig:sphaleron_cartoon}. Thus, unlike the \emph{suppressed} quantum tunneling through the barrier via instantons~\cite{tHooft:1976snw}, sphalerons are \emph{enhanced} at high temperatures~\cite{Christ:1979zm,Klinkhamer:1984di}. The name ``sphaleron'' was exactly coined to capture these features~\cite{Klinkhamer:1984di}, as it comes from the Greek word \textgreek{σφᾰλερός} (sph\v{a}ler\'{o}s), which can be translated as ``slippery, ready to fall''.

An intriguing physical observable related to sphaleron configurations is the strong \emph{sphaleron rate} $\Gamma_{\S}$, the rate of real-time (i.e., Minkowski time) sphaleron transitions in QCD. It is defined as:
\beq\label{eq:sphal_rate_def_real_time}
\Gamma_{\S} \equiv \int \dd^3 x \, \dd t \, \braket{q(\vec{x},t)q(\vec{0},0)}_T.
\eeq
Here, $q(\vec{x},t)$ is the topological charge density:
\beq
q(\vec{x},t) = \frac{1}{16\pi^2} \Tr\left[G_{\mu\nu}(\vec{x},t)\widetilde{G}^{\mu\nu}(\vec{x},t)\right],
\eeq
with $G_{\mu\nu} = \partial_\mu A_\nu - \partial_\nu A_\mu + \ii [A_\mu,A_\nu]$ and $\widetilde{G}_{\mu\nu}=\frac{1}{2}\varepsilon_{\mu\nu\rho\sigma} G^{\mu\nu}$ the gauge field strength and its dual tensor. The thermal expectation value $\braket{\dots}_T$ in Eq.~\eqref{eq:sphal_rate_def_real_time} is defined as:
\beq
\braket{q(\vec{x},t)q(\vec{0},0)}_{T} \equiv \frac{1}{Z(T)}\Tr\left\{\ee^{-\mathcal{H}_{\QCD}/T} \, q(\vec{x},t)q(\vec{0},0)\right\}, \qquad Z(T) \equiv \Tr\left\{\ee^{-\mathcal{H}_{\QCD}/T}\right\}.
\eeq
From the discussion so far, it is clear that the QCD sphaleron rate is a physical quantity  characterizing the intricate non-perturbative dynamics of strong interactions, and has thus deep theoretical significance \emph{per se}. In addition to this, however, it is also of great phenomenological relevance, both for collider physics and for Dark Matter searches.

Concerning collider physics, by virtue of the axial anomaly~\cite{tHooft:1976rip}, sphaleron transitions can create chiral imbalances~\cite{Christ:1979zm}, i.e., an excess of left-handed or right-handed fermion excitations: $N_{\Right} - N_{\Left} \ne 0$. If such an event takes place in a hot, dense and magnetized out-of-equilibrium strongly-interacting medium, such as the fireball forming for few instants after a heavy-ion collision (see, e.g., Ref.~\cite{Busza:2018rrf}), this can lead to a phenomenon known as the \emph{Chiral Magnetic Effect} (CME)~\cite{Kharzeev:2007jp,Fukushima:2008xe,Fukushima:2010vw,Kharzeev:2013ffa}.

\begin{wrapfigure}{r}{0.5\textwidth}
\centering
\includegraphics[scale=0.25]{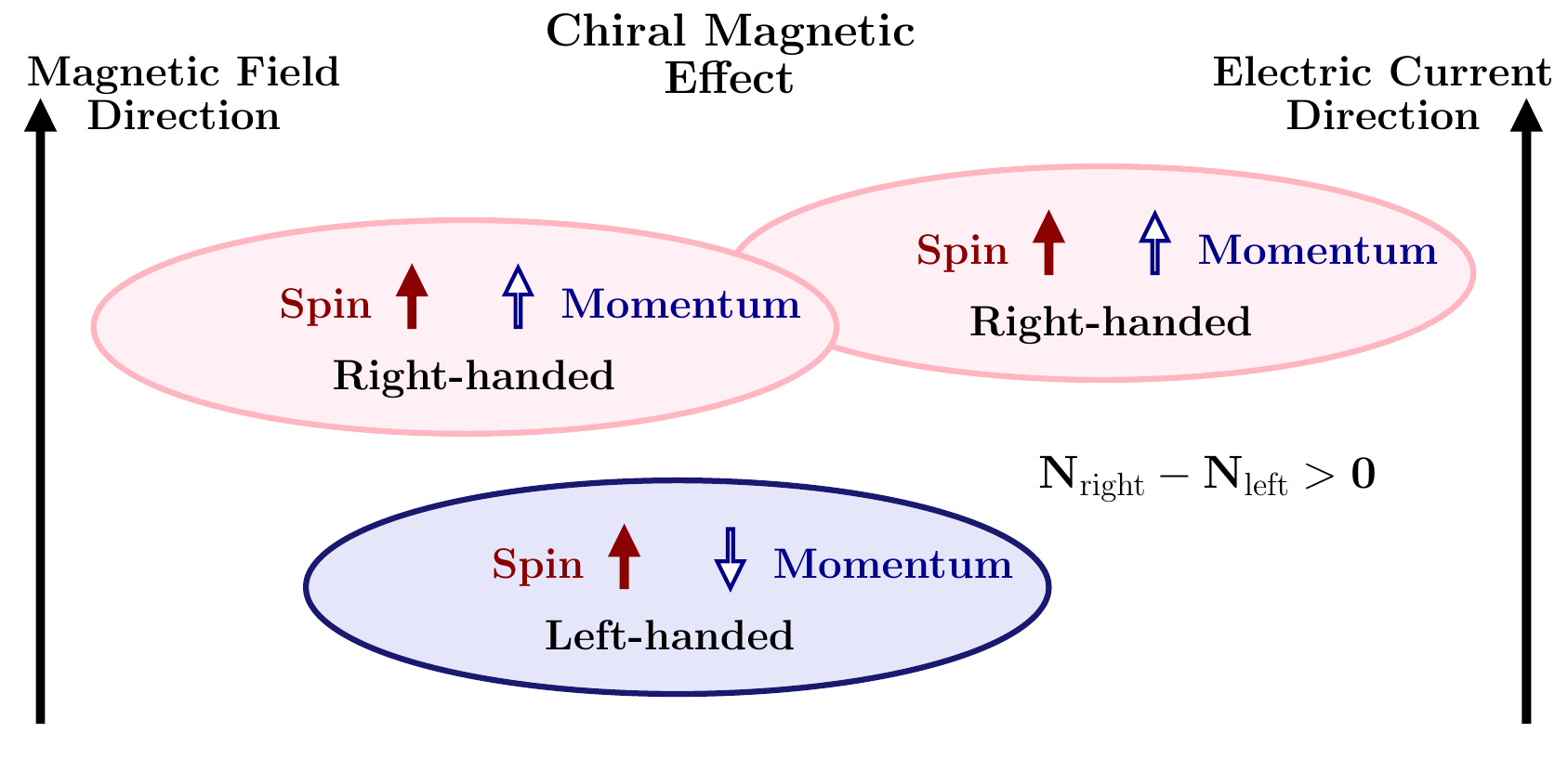}
\caption{Cartoon representing the Chiral Magnetic Effect in the presence of a sphaleron-induced quark chiral imbalance.}
\label{fig:CME_cartoon}
\end{wrapfigure}

\noindent In a few words, the coupling between the quark electric charges and the external magnetic field causes an alignment of quark spins along the magnetic field itself. When quark excitations carry also a definite chirality, this in turns causes also an alignment of their momenta along the magnetic field direction. Therefore, the presence of an imbalance between left-handed and right-handed modes, caused by a sphaleron transition, leads to a net electric current flowing in the medium in the parallel direction to the magnetic field. This is illustrated in the cartoon in Fig.~\ref{fig:CME_cartoon}. In this context, the sphaleron rate describes the relaxation of the axial quark number density $\rho_{\five}$ as the medium equilibrates in the presence of $N_{\f}^{\light}$ light quark species~\cite{Moore:2010jd,BarrosoMancha:2022mbj}:
\beq
\frac{\dd \rho_{\five}}{\dd t} = -\frac{N_{\f}^{\light}}{T^3} \Gamma_{\S} \, \rho_{\five}.
\eeq
Signatures of the CME are currently being searched in heavy-ion collision experiments, see, e.g., Ref.~\cite{Feng:2025yte}, thus theoretical inputs about $\Gamma_\S$ are of the utmost importance for this quest.

In the context of Dark Matter cosmological searches, instead, the QCD sphaleron rate $\Gamma_\S$, and its generalization to non-zero momenta (sometimes simply called the ``topological rate''):
\beq\label{eq:topological_rate_nonzeromomenta}
\Gamma(E,p) = \int \dd^3 x \, \dd t \, \ee^{\ii p^\mu x_\mu} \, \braket{q(\vec{x},t)q(\vec{0},0)}_T, \quad \Gamma_\S = \Gamma(0,0), \quad p^\mu = \left(E, \vec{p}\right), \quad p = \vert \vec{p} \vert,
\eeq
play instead a very important phenomenological role for QCD axion physics~\cite{Notari:2022ffe,Bianchini:2023ubu,OHare:2024nmr,Bouzoud:2026rur}. In the hot cosmological medium permeating the early Universe, sphaleron transitions are effective in creating hot axions by virtue of the axion-gluon interaction term $\propto a(x)q(x)$. This process is expected to have happened out of equilibrium, thus, in order to compute the time-evolution of the axion number distribution function $f_p$, one has to solve the following Boltzmann equation~\cite{Notari:2022ffe},
\beq
\frac{\dd f_p}{\dd t} = \left(1+f_p\right) \Gamma^{\pp}_{p} - f_{p} \Gamma^{\mm}_{p},
\eeq
where the axion creation/annihilation rates $\Gamma^{\pp}_p/\Gamma^{\mm}_p$ are given by:
\beq
\Gamma^{\mm}_p = \ee^{\frac{E(p)}{T}} \Gamma^{\pp}_p = \frac{1}{2E(p)f_a} \Gamma(E(p),p), \qquad E^2(p) = m_a^2 + p^2,
\eeq
with $m_a$ and $f_a$ the axion mass and decay constant respectively. Theoretical estimates to compute the abundance of relic axions, which can in principle be measured by upcoming cosmological surveys, require the non-perturbative QCD topological rate as a fundamental input, especially for temperatures around and above the QCD chiral crossover, where non-perturbative effects are important.

Due to its theoretical and phenomenological importance, and by virtue of its intrinsic non-perturbative nature (especially at the temperatures of interest for physical applications), Lattice QCD seems a natural tool to compute the QCD sphaleron rate from first principles. However, being Monte Carlo calculations of lattice-regularized Quantum Field Theory based on the Euclidean formulation, this is an highly non-trivial task due to the inherent real-time nature of the sphaleron rate. More precisely, using analytic continuation, one can only relate the sphaleron rate to an Euclidean correlation function via a convolutional integral relation. Let us introduce the Euclidean time correlator of the topological charge density:
\beq
G_{\E}(\tau) = \int \dd^3 x \braket{q(\vec{x},\tau)q(\vec{0},0}, \qquad \quad \int \dd^4 x \, q(x) = Q \in \mathbb{Z},
\eeq
with $Q$ the integer-valued topological charge, $\tau$ the Euclidean time separation, and where $\braket{\dots}$ stands for the Euclidean path integral expectation value computed in the presence of a compactified temporal direction with length $1/T$. Introducing the thermal spectral density $\rho(\omega)$ of $G_\E(\tau)$,
\beq\label{eq:def_spec_dens}
G_{\E}(\tau) = - \int_0^{\infty} \frac{\dd \omega}{\pi} \rho(\omega) \frac{\cosh\left[\omega \left(\tau - \frac{1}{2T}\right)\right]}{\sinh\left(\frac{\omega}{2T}\right)} \equiv - \int_0^{\infty} \frac{\dd \omega}{\pi} \rho(\omega) K(\omega,\tau),
\eeq
the following relation holds~\cite{Meyer:2011gj,Lowdon:2022keu}:
\beq\label{eq:Kubo_eq_sphalrate}
\Gamma_\S = 2T \lim_{\omega \,\to\, 0} \frac{\rho(\omega)}{\omega}.
\eeq
In practice, extracting the sphaleron rate from a lattice calculation requires the calculation of $G_{\E}(\tau)$, and then the inversion of Eq.~\eqref{eq:def_spec_dens} to extract the spectral density [related to $\Gamma_\S$ via Eq.~\eqref{eq:Kubo_eq_sphalrate}]. This task falls under a broad class of mathematical problems known as \emph{inverse problems}. Inverse problems are notoriously difficult to solve numerically since they are ill-posed, i.e., even a small fluctuation of the input data can cause a huge change in the output results. Given that lattice-computed correlation functions are affected by statistical uncertainties and are only known for a discrete set of temporal separations, statistical errors on the spectral density will blow up if no strategy is employed to improve the conditioning of the inverse problem. This explains why, until a few of years ago, only a few preliminary lattice attempts to compute $\Gamma_{\S}$, all limited to the SU(3) pure-gauge theory, have been performed~\cite{Kotov:2018aaa,Altenkort:2020axj,BarrosoMancha:2022mbj}.

In our recent paper~\cite{Bonanno:2023thi}, we significantly advanced the state of the art, providing the first non-perturbative investigation of $\Gamma_\S$ in $N_\f=2+1$ QCD with physical quark masses in the temperature range $1.5 \lesssim T/T_c\lesssim 4$, with $T_c\simeq155$ MeV the QCD chiral crossover temperature. This advancement was possible thanks to the novel methodology we introduced in Ref.~\cite{Bonanno:2023ljc}. In recent years, many studies addressed possible numerical procedures to solve inverse problems in Lattice Field Theory~\cite{Astrakhantsev:2018oue,Hansen:2019idp,Boito:2022njs,Rothkopf:2022fyo,Altenkort:2023oms,Bruno:2024fqc,DelDebbio:2024lwm}. The method of Ref.~\cite{Bonanno:2023ljc} exploits one of these new proposals, the Hansen--Lupo--Tantalo (HLT) method~\cite{Hansen:2019idp} --- based on a non-trivial modification of the Backus--Gilbert method~\cite{BackusGilbert1968:aaa} --- which has been extensively employed in recent years~\cite{ExtendedTwistedMassCollaborationETMC:2022sta,DelDebbio:2022qgu,Frezzotti:2023nun,Evangelista:2023fmt,ExtendedTwistedMass:2024myu,Bennett:2024cqv,DeSantis:2025qbb,DeSantis:2025yfm,Frezzotti:2025hif, TELOS:2025ash}.

This manuscript is organized as follows. The new method of Ref.~\cite{Bonanno:2023ljc} is presented in Sec.~\ref{sec:method}. The results of Ref.~\cite{Bonanno:2023thi} about the temperature dependence of the QCD sphaleron rate are discussed in Sec.~\ref{sec:results}. Finally, in Sec.~\ref{sec:conclusions}, I will draw my conclusions and discuss future outlooks, and how algorithmic improvements of Ref.~\cite{Bonanno:2024zyn} will be key to achieve them.

\section{Non-perturbative strong sphaleron rate from Lattice QCD: a novel approach}\label{sec:method}

\noindent This section summarizes the method of~\cite{Bonanno:2023ljc} to compute the sphaleron rate from Lattice QCD.

\subsection{The inverse problem resolution}

Let us rewrite the target inverse problem as:
\beq\label{eq:inverse_problem_sphalrate_HLT}
G_{\L}(\tau) = - \int_0^{\infty} \frac{\dd \omega}{\pi} \frac{\rho_\L(\omega)}{\omega} K^\prime(\omega,\tau), \qquad K^\prime(\omega,\tau) = \omega K(\omega,\tau),
\eeq
where $G_{\L}(\tau)$ is the lattice topological charge density correlator (whose calculation will be described in the next section~\ref{sec:calc_TCDC}), and $\rho_\L(\omega)$ its spectral density. Let us look for a solution for the spectral density of the form:
\beq\label{eq:specdens_HLT_sol}
\frac{\rhobar_\L(\omegabar)}{\omegabar} = - \pi \sum_{\tau \, = \, 0}^{1/T} \, g_\tau(\omegabar) \, G_{\L}(\tau),
\eeq
with $g_\tau(\omegabar)$ unknown coefficients to be determined, and where the sum runs over the discretized times $\tau=an_t$, $n_t=0,1,\dots,N_t$, with $a$ the lattice spacing and $N_t=1/(aT)$ the number of lattice temporal sites. This implies that the lattice sphaleron rate $\Gamma_\L$ is given by:
\beq\label{eq:sphalrate_HLT_sol}
\frac{1}{2T}\Gamma_\L = \frac{\rhobar_\L(\omegabar)}{\omegabar} \bigg\vert_{\omegabar \, = \, 0} = - \pi \sum_{\tau \, = \, 0}^{1/T}  g_\tau \, G_{\L}(\tau),
\eeq
where $g_\tau$ is a short-hand for $g_\tau(\omegabar=0)$. Combining Eq.~\eqref{eq:specdens_HLT_sol} and Eq.~\eqref{eq:inverse_problem_sphalrate_HLT}, one obtains the following consistency relation:
\beq
\frac{\rhobar_\L(\omegabar)}{\omegabar} = \int_0^{\infty} \dd\omega \, \Delta(\omega,\omegabar) \frac{\rho_\L(\omega)}{\omega}.
\eeq
This is a \emph{smearing relation} between the actual spectral density $\rho_\L(\omega)/\omega$ and the solution $\rhobar_\L(\omegabar)/\omegabar$. The smearing kernel $\Delta(\omega,\omegabar)$, also known as \emph{resolution function}, is given by:
\beq
\Delta(\omega,\omegabar) = \sum_{\tau \, = \, 0}^{1/T} \, g_\tau(\omegabar) \, K^\prime(\omega,\tau).
\eeq
Determining the sphaleron rate has thus been rephrased into the problem of determining the coefficients $g_\tau$ in such a way that $\Delta(\omega) \equiv \Delta(\omega,\omegabar=0)$ is sufficiently peaked around $\omegabar=0$, so that $\rhobar_\L(\omegabar)/\omegabar$ is a good approximation of $\rho_\L(\omega)/\omega$ around the origin.

The HLT method exactly allows to determine the $g_\tau$ coefficients from the minimization of the following functional:
\beq
F_\lambda[g] = (1-\lambda) A_2[g] + \lambda B[g],
\eeq
with
\beq
A_\alpha[g] = \int_0^{\infty} \dd \omega \, \vert \Delta(\omega) - \delta_\sigma(\omega) \vert^2 \, \ee^{\alpha a\omega}, \qquad
B[g] = \frac{1}{G^2_{\L}(0)} \sum_{\tau_1, \tau_2} \mathrm{Cov}_{\tau_1\tau_2} g_{\tau_1} g_{\tau_2}.
\eeq
The functional $A_2[g]$ expresses the difference between the smearing kernel $\Delta(\omega)$ and a given target smearing kernel $\delta_\sigma(\omega)$, which can be chosen with a certain degree of arbitrariness. In~\cite{Bonanno:2023ljc,Bonanno:2023thi}, inspired by the functional form of $K^\prime(\omega,\tau)$, we chose:
\beq
\delta_\sigma(\omega) = \left(\frac{2}{\pi\sigma}\right)^2 \frac{\omega}{\sinh\left(\frac{\omega}{2T}\right)}, \quad \text{ which satisfies } \quad \lim_{\sigma \, \to \, 0} \delta_\sigma(\omega) = \delta(\omega).
\eeq
If the smearing kernel $\Delta(\omega)$ is close enough to the target one $\delta_\sigma(\omega)$, its width around the desired point (the origin, in the case at hand) will be given by $\sigma$, which is an input of the HLT method. Being able to control $\sigma$ is a great advantage of HLT over the original Backus--Gilbert, both because it allows to control possible systematic effects introduced by the smearing, and because it allows to keep it constant as other parameters (e.g., the lattice spacing) are varied. The functional $B[g]$, instead, is directly proportional to the statistical uncertainties on the lattice correlator $G_{\L}(\tau)$ via its covariance matrix $\mathrm{Cov}_{\tau_1\tau_2}$, and acts as a regulator of the inverse problem.

The computation of the sphaleron rate requires to minimize $F_\lambda[g]$ for several values of $\lambda$ in order to look for the optimal trade-off between $A_2[g]$ and $B[g]$, i.e., between statistical and systematic uncertainties. Indeed, when $\lambda\to 0$ and $A_2[g]$ dominates, $\Delta(\omega)$ will be very close to $\delta_\sigma(\omega)$, but fluctuations on $g_\tau$ will explode due to the inverse problem being ill-posed. As $\lambda \to 1$, $B[g]$ will dominate, and the inverse problem will be regularized. This will damp the fluctuations of $g_\tau$ at the price of introducing a systematic error, due to the fact that the shape of $\Delta(\omega)$ is unconstrained and the smearing procedure is out of control. The idea is therefore to vary $\lambda$ and look for an intermediate regime where a plateau in the resulting values of $\Gamma_\L$ is observed. When such plateau is found before statistical errors explode, it signals that systematic errors are under control, and the plateau can be used to assess the final value and uncertainty of the sphaleron rate.

\begin{figure}[!t]
\centering
\includegraphics[scale=0.385]{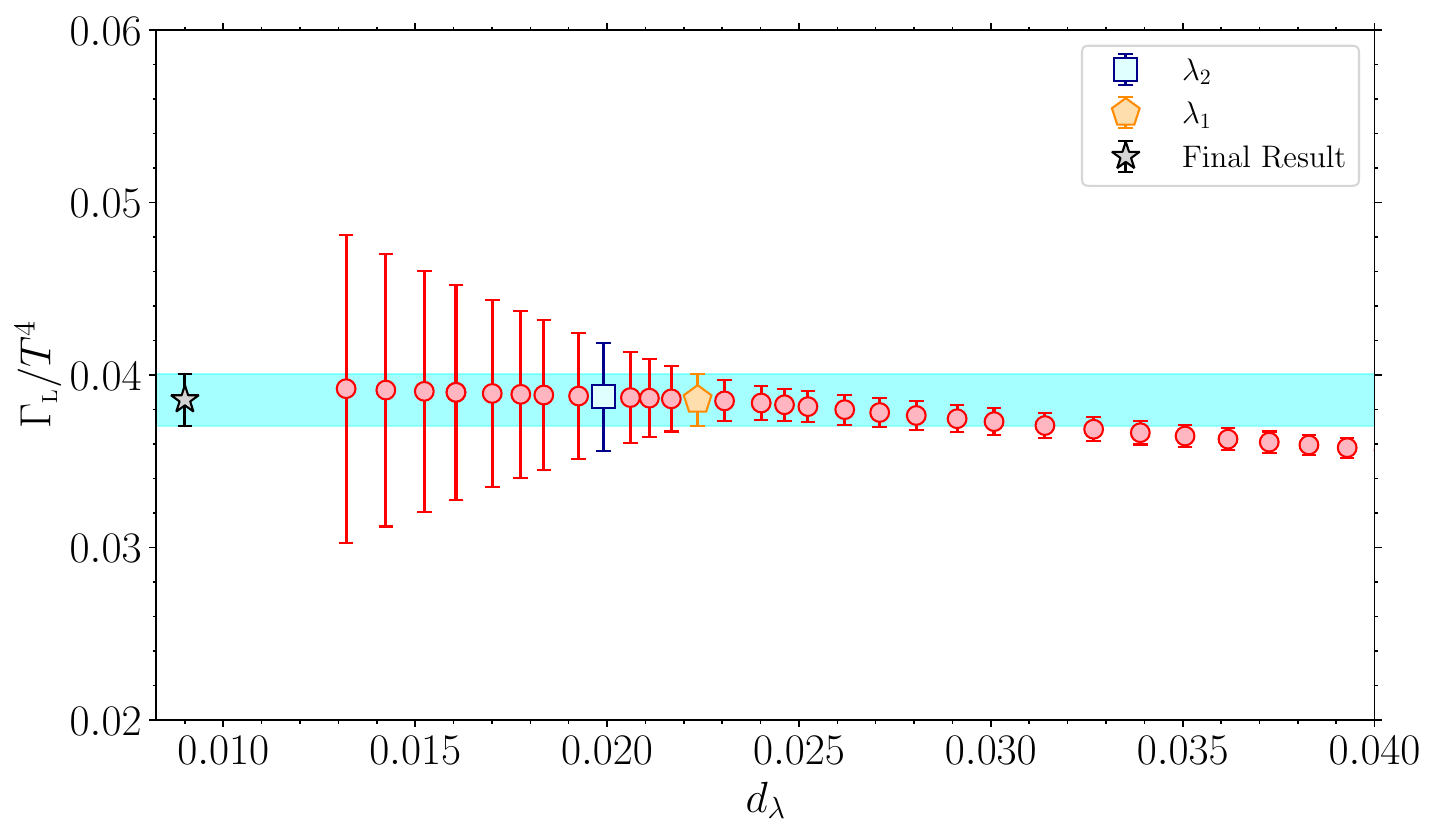}
\includegraphics[scale=0.385]{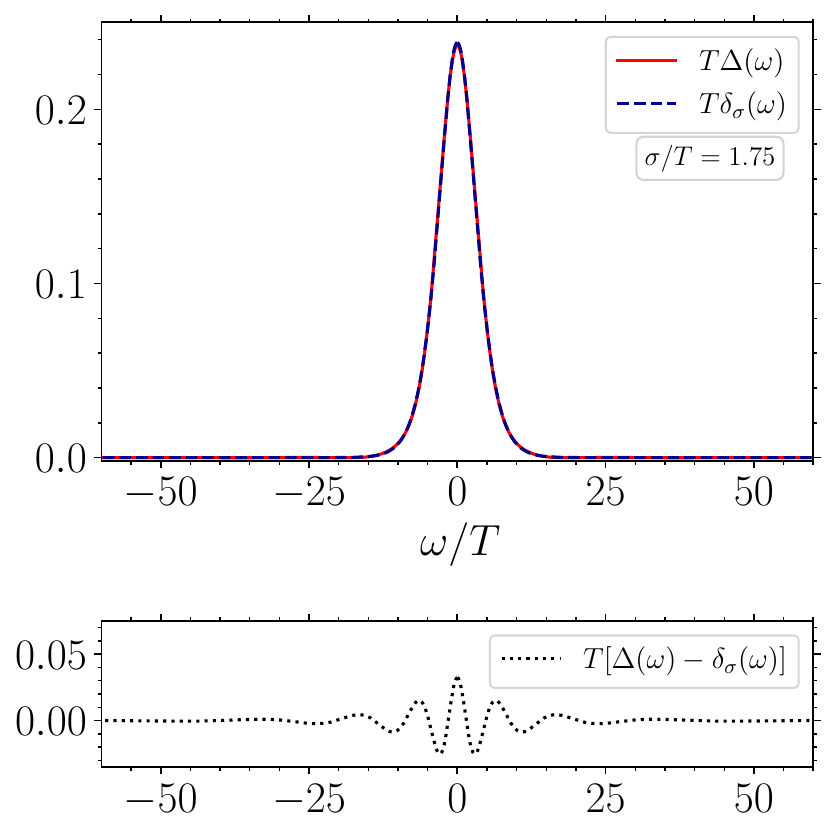}
\caption{Left panel: example of the calculation of the sphaleron rate from the inverse problem resolution via the HLT method. Right panel: comparison between the reconstructed smearing kernel $\Delta(\omega)$ for $\lambda=\lambda_1$ and the chosen target one $\delta_\sigma(\omega)$. Figures from Ref.~\cite{Bonanno:2023ljc}.}
\label{fig:HLT_example}
\end{figure}

An example of this procedure is illustrated in the left panel of Fig.~\ref{fig:HLT_example}, while the right panel shows an example of the reconstructed smearing kernel versus the target one. Since $\lambda$ typically varies across several orders of magnitude, it is better to look for a plateau in terms of the figure of merit:
\beq
d_\lambda = \sqrt{\frac{A_0[g_\tau^*(\lambda)]}{A_0[0]}},
\eeq
where $g_\tau^*(\lambda)$ is the minimum of $F_\lambda[g]$. The quantity $d_\lambda$ expresses the distance between the smearing kernel $\Delta(\omega)$ and the target smearing kernel $\delta_\sigma(\omega)$ for a given $\lambda$, and is thus a good figure of merit to quantify the reliability of the numerical solution of the inverse problem. Indeed, if $d_\lambda$ is small, the reconstructed kernel is close to the target one, and one has control over the smearing of the spectral density. Once a plateau for $\Gamma_\L$ as a function of $d_{\lambda}$ is identified, we select two values $\lambda_1>\lambda_2$. The value of the rate $\Gamma_1$ corresponding to $\lambda_1$ gives the central value and the statistical uncertainty of the final result. The value $\Gamma_2$ corresponding to $\lambda_2$ is used to assess the systematic error (which is summed in quadrature to the statistical one) via:
\beq
\mathrm{Err}_{\rm syst} = \big\vert \Gamma_1 - \Gamma_2 \big\vert \, \mathrm{erf}\left( \frac{1}{\sqrt{2}}\frac{\big\vert \Gamma_1 - \Gamma_2 \big\vert}{\sqrt{\mathrm{Err}^2_{\mathrm{stat},1}+\mathrm{Err}^2_{\mathrm{stat},2}}}\right).
\eeq

\subsection{Determination of the Euclidean topological charge density correlator}\label{sec:calc_TCDC}

The determination of the Euclidean topological charge density correlator from the lattice is something that is challenging by itself. It is well known that gluonic lattice topological charge formulations, such as the customary lattice clover definition $Q_{\clov}$,
\beq
Q_{\clov} &=& \sum_x q_{\clov}(x) = \frac{1}{32\pi^2} \sum_x \sum_{\mu\nu\rho\sigma}\varepsilon_{\mu\nu\rho\sigma}\Tr\left[C_{\mu\nu}(x)C_{\rho\sigma}(x)\right],\\
C_{\mu\nu}(x) &=& \frac{1}{4} \Im\left\{U_{\mu\nu}(x) + U_{-\nu, \mu}(x) + U_{\nu, -\mu}(x) + U_{-\mu, -\nu}(x)\right\},\\
U_{\mu\nu}(x) &=& U_\mu(x) U_\nu(x+a\hat{\mu}) U^\dagger_\mu(x+a\hat{\nu}) U^\dagger_\nu(x), \quad U_{-\mu}(x) = U^\dagger_{\mu}(x-a\hat{\mu}),
\eeq
suffers for multiplicative renormalizations due to ultra-violet fluctuations at the scale of the lattice spacing. Moreover, further divergent additive renormalizations appear when computing integrals of correlation functions of $q(x)$, such as the topological susceptibility $\chi=\braket{Q^2}/V=\int\dd^4x\,\braket{q(x)q(0)}=\int\dd\tau\,G_{\E}(\tau)$. To deal with these renormalizations, it is customary to compute lattice gluonic topological charge definitions on smoothened gauge configurations. Thus, the lattice topological charge density correlator is computed on smoothened configurations too.

\begin{wrapfigure}{r}{0.5\textwidth}
\centering
\includegraphics[scale=0.31]{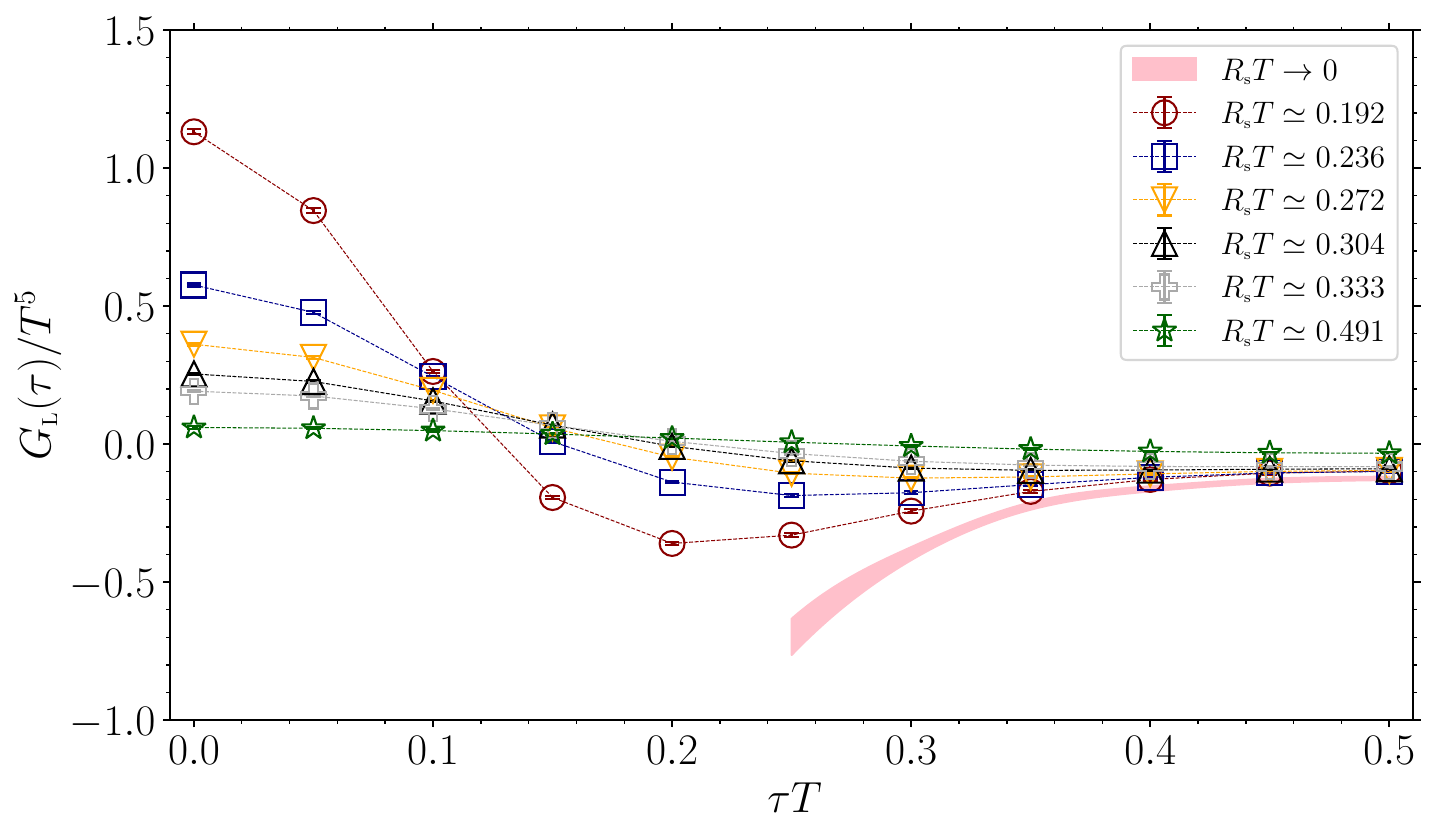}
\caption{Comparison of determinations of $G_\L(\tau)$ at finite lattice spacing for a few values of $\Rs$ with the $\Rs\to0$-extrapolated continuum correlator $G_\E(\tau)$. Figure from Ref.~\cite{Bonanno:2023ljc}.}
\label{fig:charge_corr_example}
\end{wrapfigure}

Smoothing introduces a new length scale in the game, the \emph{smoothing radius} $\Rs$, the scale below which fluctuations are damped. This scale is proportional to the square root of the amount of smoothing performed. Smoothing unavoidably modifies the short-distance behavior of the lattice correlator $G_\L(\tau)$. This is clearly illustrated in Fig.~\ref{fig:charge_corr_example}. General theoretical arguments based on reflection positivity imply that the continuum correlator $G_\E(\tau)$ must be negative for every positive time separation $\tau>0$. On the lattice, however, due to smoothing, this condition is violated when $\tau < \Rs$. In this regime, the lattice correlator $G_\L(\tau)$ becomes positive and unphysical. The negativity of the correlator is only recovered in the limit $\Rs \to 0$, which must be taken after the continuum limit $a\to 0$. This is shown in Fig.~\ref{fig:charge_corr_example}, which displays the negativity of the continuum correlator $G_{\E}(\tau)$, obtained after the double extrapolation $a\to 0$ and $\Rs\to 0$.

In our studies~\cite{Bonanno:2023ljc,Bonanno:2023thi} we employed cooling~\cite{Berg:1981nw, Iwasaki:1983bv, Teper:1985rb, Ilgenfritz:1985dz,Campostrini:1989dh} as our smoothing method of choice. In this case, the square smoothing radius is proportional to the number of cooling steps:
\beq
\frac{\Rs}{a} = \sqrt{\frac{8}{3}n_{\cool}}.
\eeq
Other smoothing methods, such as gradient flow~\cite{Narayanan:2006rf,Luscher:2009eq,Luscher:2010iy} or stout smearing~\cite{Morningstar:2003gk}, have been shown to give consistent results once the smoothing radii are matched to one another~\cite{Bonati:2014tqa} (see also~\cite{Alexandrou:2017hqw}).

\section{Numerical results for the QCD sphaleron rate}\label{sec:results}

\noindent This section summarizes the main results of~\cite{Bonanno:2023thi} about the QCD sphaleron rate.

\subsection{Controlling finite-\texorpdfstring{$a$}{a}, finite-\texorpdfstring{$\Rs$}{Rs}, and finite-\texorpdfstring{$\sigma$}{sigma} effects on the sphaleron rate}

Using the methods described in Sec.~\ref{sec:method}, in Ref.~\cite{Bonanno:2023thi} we determined the sphaleron rate for 5 temperatures in the range $1.5 \lesssim T/T_c \lesssim 4$ in full QCD with $N_\f=2+1$ dynamical flavors, corresponding to a light quark doublet with mass $m_\ell=m_{\rm u}=m_{\rm d}$ and a strange quark with mass $m_{\rm s}$, both tuned at the physical point. More precisely, we tuned the bare parameters to have $m_\pi = m_\pi^{\phys} \simeq 135$ MeV and $R=m_{\ell}/m_{\rm s}=R^{\phys} \simeq 0.036$. Our simulations are performed at fixed spatial lattice size $LT=4$, which is expected to be sufficiently large to avoid significant finite-volume effects. Due to the strong suppression of the topological susceptibility $\chi\sim1/T^8$ in the high-temperature phase of QCD, topological charge fluctuations become very rare since $\braket{Q^2}=\chi V = \chi L^3/T \sim 1/T^{12}\ll 1$. This introduces a sampling problem in the topological charge, which in~\cite{Bonanno:2023thi} we overcame by adopting a multicanonical algorithmic approach~\cite{Bonati:2017woi,Jahn:2018dke,Bonati:2018blm}.

\begin{figure}[!t]
\centering
\includegraphics[scale=0.3]{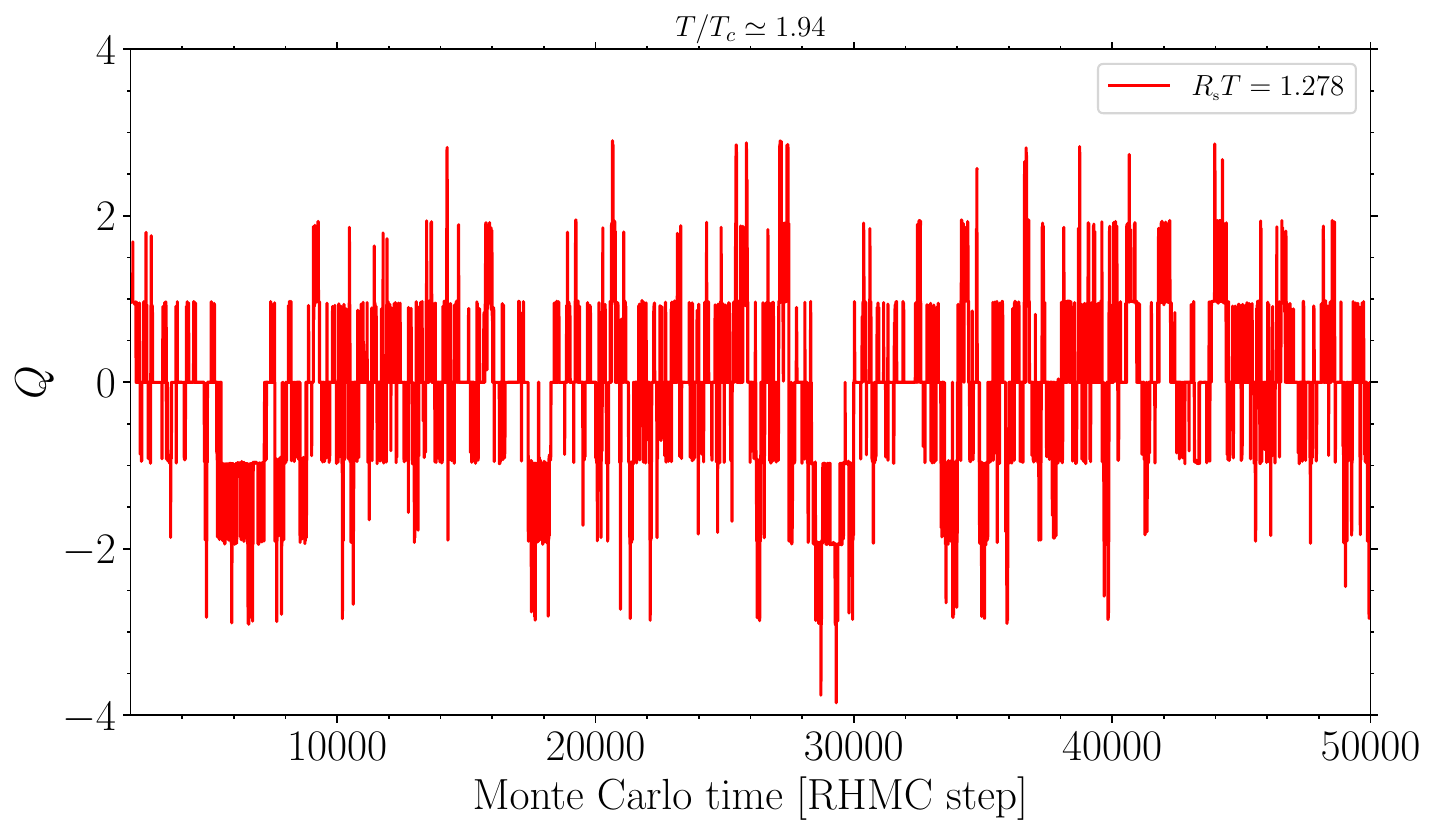}
\includegraphics[scale=0.3]{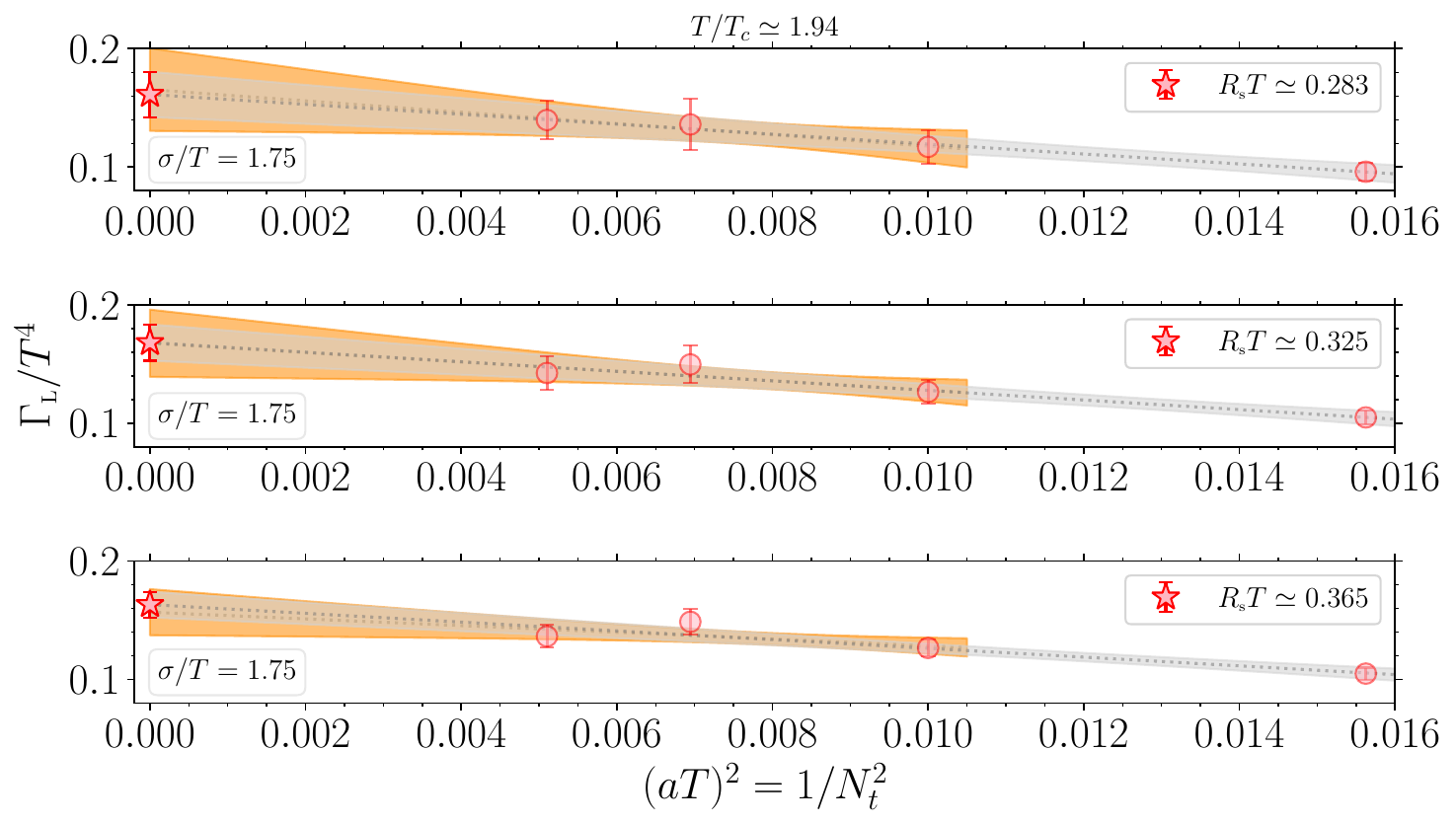}
\includegraphics[scale=0.3]{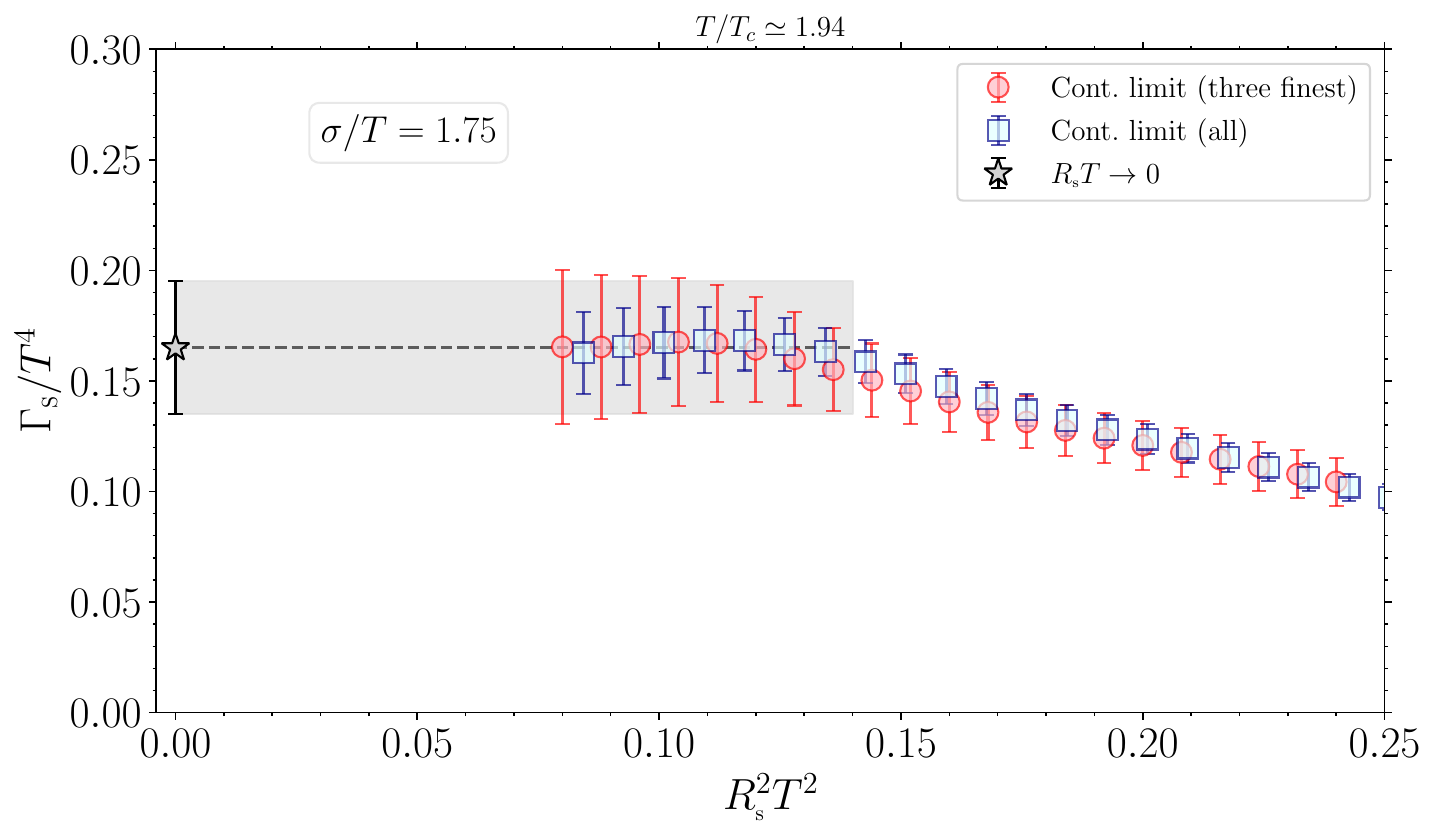}
\includegraphics[scale=0.3]{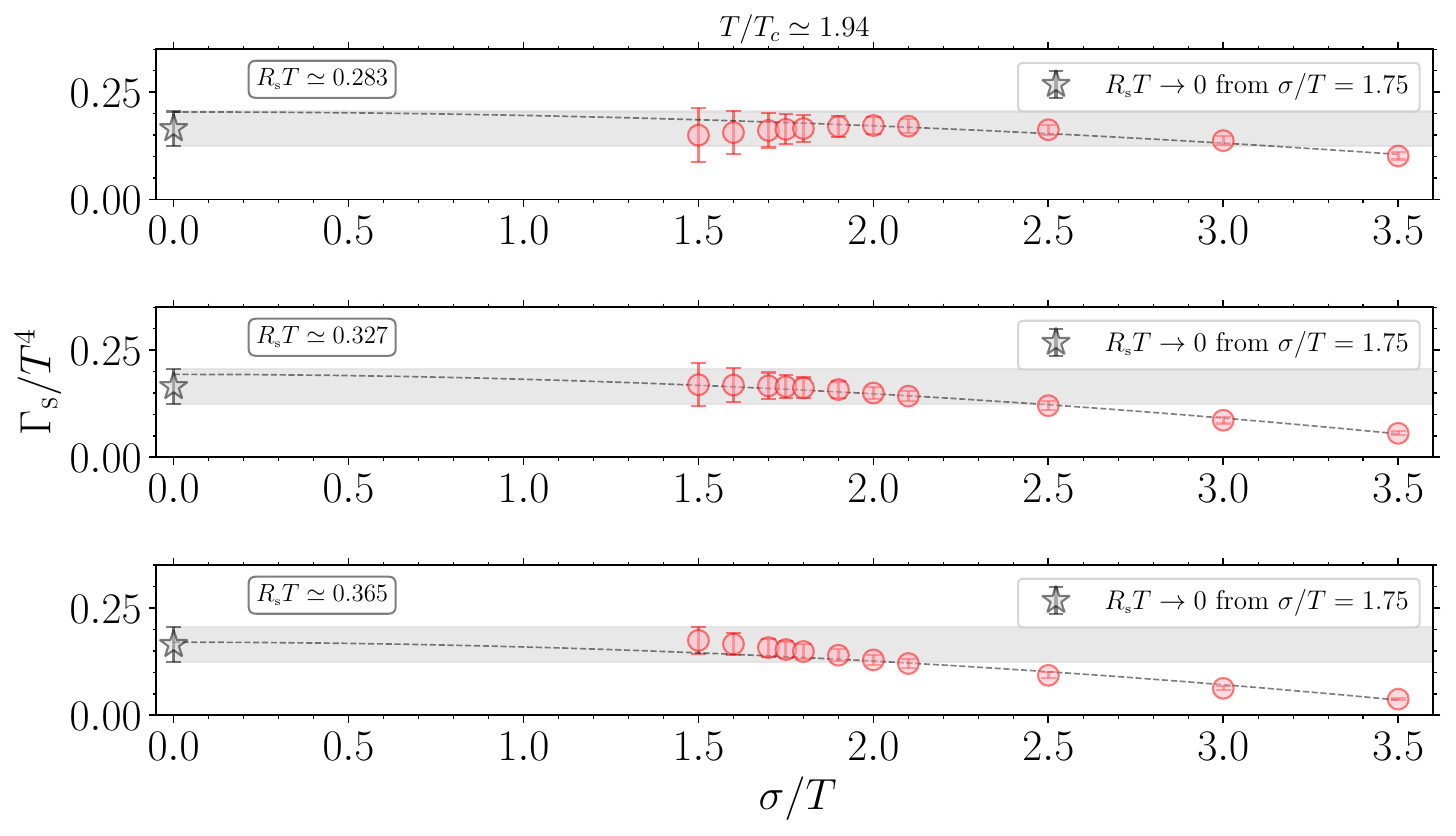}
\caption{Example of the calculation of the sphaleron rate for $T=300~\mathrm{MeV} \simeq 1.94 T_c$. Top left panel: Monte Carlo evolution of the cooled clover lattice topological charge for the finest lattice spacing explored at this temperature, $aT=1/14$. Top right panel: continuum limit of $\Gamma_\L/T^4$ at fixed $\sigma/T=1.75$ and $\Rs T$ for the three smallest smoothing radii explored. Bottom left panel: smoothing-radius dependence of the continuum results for $\Gamma_\S/T^4$ (fixed $\sigma/T=1.75$). Bottom right panel: smearing-width dependence of the continuum results for $\Gamma_\S/T^4$ at fixed $\Rs/T$ for the three smallest smoothing radii explored. Figures from Ref.~\cite{Bonanno:2023thi}.}
\label{fig:sphal_rate_T300_example}
\end{figure}

The sphaleron rate $\Gamma_\L$ is calculated for several lattice spacings $aT=1/N_t$, for several values of the smoothing radius $\Rs T$ used to compute the lattice topological charge density correlator $G_\L(\tau)$, and for several values of the smearing width $\sigma/T$ used to solve the inverse problem via the HLT method. The final result was obtained in~\cite{Bonanno:2023thi} as follows. First, we took the continuum limit $aT\to 0$ at fixed $\Rs T$ and $\sigma/T$. Then, we took the limits $\Rs T \to 0$ and $\sigma/T\to 0$ of the continuum results.\\
In Fig.~\ref{fig:sphal_rate_T300_example}, the procedure to compute $\Gamma_\S$ is exemplified for $T\simeq 300$ MeV, i.e., $T/T_c\simeq 1.94$. The top left panel shows the Monte Carlo history of the topological charge. As it can be seen, our multicanonical simulations completely avoid the sampling problem of the topological charge due to the suppression of $\braket{Q^2}$ at high temperatures. The top right panel shows the continuum limit of $\Gamma_\L/T^4$ for the 3 smallest values of $\Rs T$ explored. All shown continuum limits are taken at fixed $\sigma/T=1.75$. They all turn out to be rather smooth, and no significant difference is observed if the coarsest lattice spacing is included/excluded in the extrapolation. The bottom left panel shows the $\Rs$-dependence of the continuum extrapolations of $\Gamma_\S/T^4$ for fixed $\sigma/T=1.75$. As it can be seen, below $\Rs T\simeq 0.375$ the sphaleron rate exhibits a plateau as a function of the smoothing radius. The value of such plateau is taken as our $\Rs T \to 0$ limit. This behavior has a clear physical explanation: the smoothing radius $\Rs$ is a ultra-violet scale, while the sphaleron rate is the zero-frequency limit of the spectral density slope, which receives the dominant contributions from the large-$\tau$ tail (i.e., $\tau T$ close to 0.5) of the topological charge density correlator. Thus, when $\Rs$ is sufficiently small, there is an effective separation between these two scales, and $\Gamma_\S$ becomes independent of the smoothing radius. This is similar to what happens to the topological susceptibility, which also becomes independent of the smoothing radius in the continuum limit. Finally, the bottom right panel shows the $\sigma/T$-dependence of the continuum determinations of $\Gamma_\S/T^4$ for the 3 smallest values of $\Rs T$ explored (thus all lying in the plateau previously identified). The general theoretical expectation for an even smearing kernel $\Delta(\omega)=\Delta(-\omega)$ like ours is that~\cite{Frezzotti:2023nun}:
\beq\label{eq:sphlrate_sigma_dep}
\Gamma_\S(\sigma) = \Gamma_\S(0) \left[ 1 + \mathcal{O}(\sigma^2) \right].
\eeq
The results shown in Fig.~\ref{fig:sphal_rate_T300_example} are in agreement with this prediction. Indeed, below $\sigma/T \simeq 2$ one observes that the sphaleron rate becomes rather insensitive to the smearing width, signaling that the smeared spectral density computed from HLT faithfully represents the physical one in this regime, and that $\mathcal{O}(\sigma^2)$ corrections become smaller than our statistical uncertainties. This result is further corroborated by the fact that explicit $\sigma/T\to 0$ extrapolations according to Eq.~\eqref{eq:sphlrate_sigma_dep}, involving also results obtained for $\sigma/T>2$, all turn out to be compatible with the value at the plateau observed in the bottom left panel of Fig.~\ref{fig:sphal_rate_T300_example} for $\sigma/T=1.75$. Therefore, that value is taken to be our final result for $\Gamma_\S/T^4$.

\subsection{The temperature-dependence of the QCD sphaleron rate}

\begin{figure}[!t]
\centering
\includegraphics[scale=0.3]{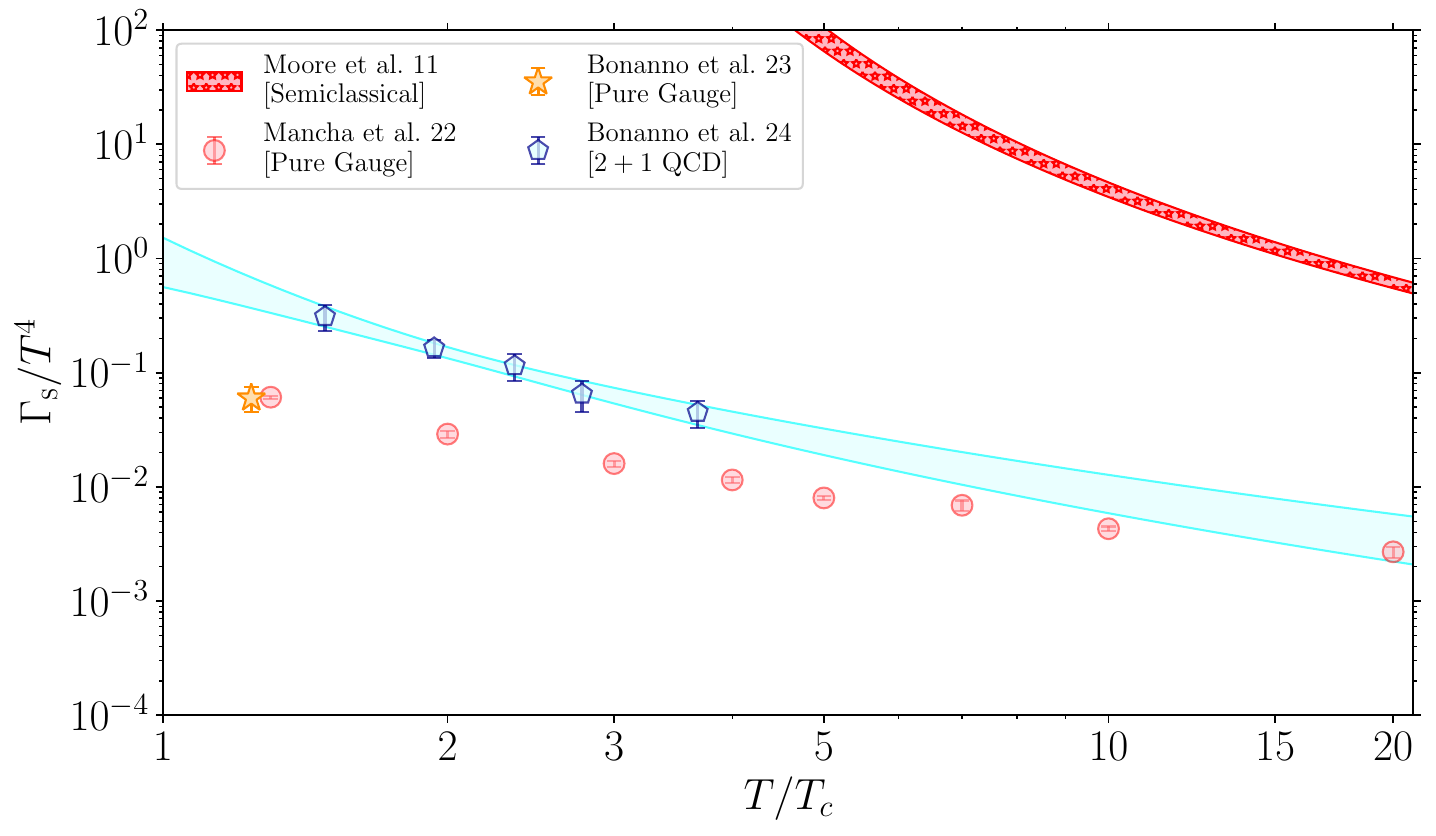}
\includegraphics[scale=0.3]{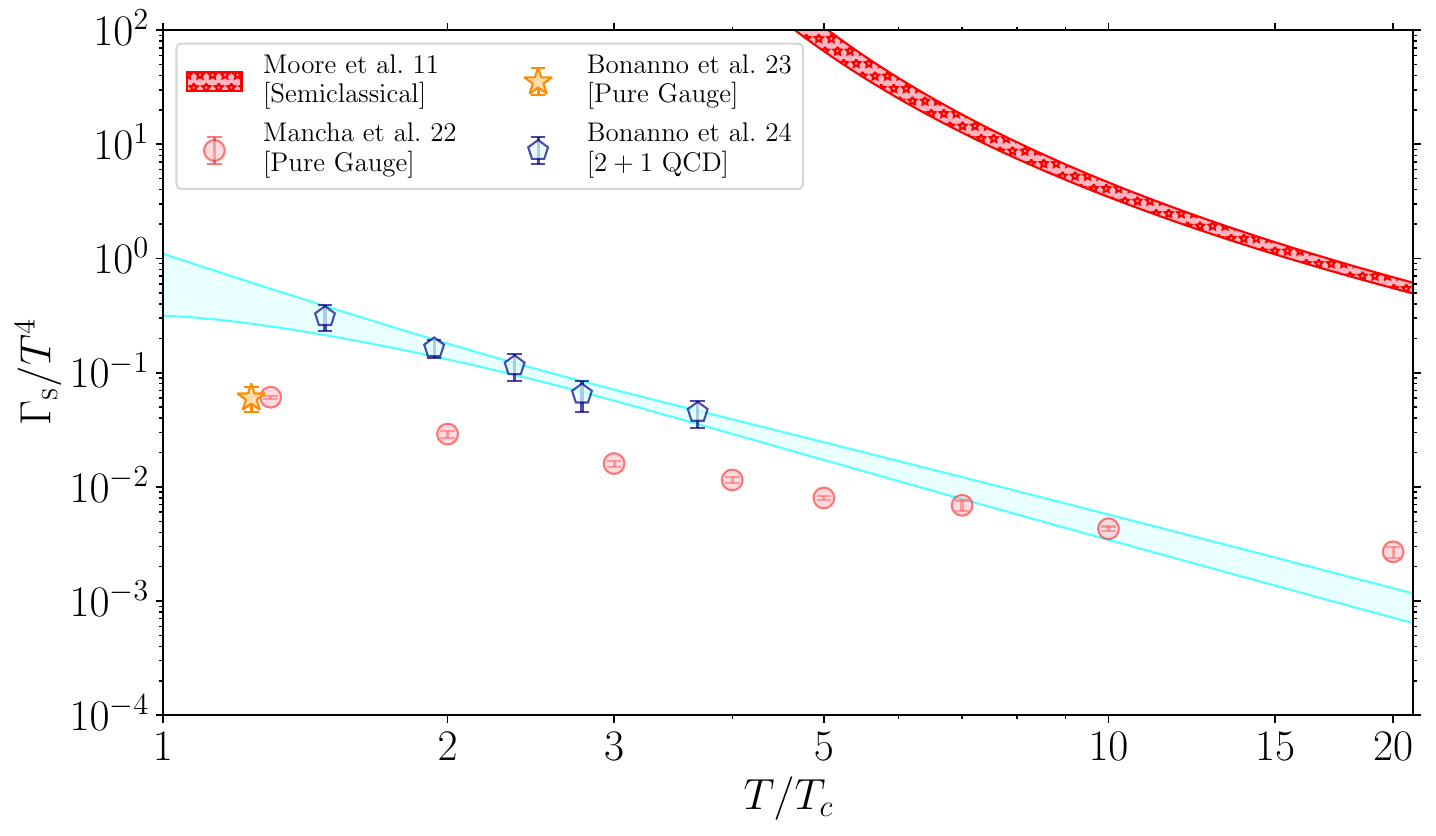}
\caption{Temperature dependence of the sphaleron rate in 2+1 QCD~\cite{Bonanno:2023thi} compared to the temperature dependence in the pure gauge theory~\cite{BarrosoMancha:2022mbj}, and to the one expected from semiclassics~\cite{Moore:2010jd}. The blue bands represent, respectively, the best fit of the full QCD data to the ansatz in Eq.~\eqref{eq:sphal_rate_ans1} (left panel), and to the ansatz in Eq.~\eqref{eq:sphal_rate_ans2} (right panel). The starred point stands for the result obtained in~\cite{Bonanno:2023ljc} in the pure gauge theory, using the same method employed in full QCD. It is a non-trivial cross-check, as it agrees with the result of~\cite{BarrosoMancha:2022mbj}, obtained without resorting to any inverse problem resolution, at a close-by temperature.  Figures from Ref.~\cite{Bonanno:2023thi}.}
\label{fig:sphalrate_vs_T}
\end{figure}

The temperature-dependence of the QCD sphaleron rate is shown in Fig.~\ref{fig:sphalrate_vs_T}, where we compare it with pure-gauge results of~\cite{BarrosoMancha:2022mbj}, and with the semiclassical computation of~\cite{Moore:2010jd}. The comparison is performed in terms of $T/T_c$, where for the pure-gauge theory the value of the critical deconfinement temperature $T_c\simeq 287$ MeV was used. Interestingly, unlike what happens with the topological susceptibility, the sphaleron rate is not suppressed by the presence of light dynamical quarks. This is likely due to the fact that $\Gamma_\S$ receives dominant contribution from the large-$\tau$ tail of $G_{\E}(\tau)$, which is not expected to be suppressed by light dynamical fermions. Our new method~\cite{Bonanno:2023ljc} was cross-checked in the pure gauge theory against the result of~\cite{BarrosoMancha:2022mbj} for a near-by temperature, finding perfect agreement. This is highly non-trivial, given that the results of~\cite{BarrosoMancha:2022mbj} were obtained with a completely different method, not relying on the inverse problem resolution.

In Ref.~\cite{Bonanno:2023thi}, we tried two different Ans\"atze to fit the $T$-dependence of $\Gamma_\S/T^4$. One is rooted on the semiclassical result of~\cite{Moore:2010jd}:
\beq
\Gamma_\S \propto \alpha^5_{\rm s}(T),
\eeq
and in the 1-loop running of $\alpha_{\rm s}(\mu) \propto 1/\log(\mu/\Lambda_{\QCD})$. In particular, we employed the following fit function:
\beq\label{eq:sphal_rate_ans1}
\frac{\Gamma_\S}{T^4} = \left(\frac{A}{\log x + B}\right)^C, \qquad x=T/T_c.
\eeq
Fixing $C=5$ (as per the semiclassical prediction) gives an excellent description of the data, shown in the left panel of Fig.~\ref{fig:sphalrate_vs_T} as a shaded band. However, leaving $C$ as a free parameter gives $C\simeq 5.6$ with a $100\%$ uncertainty. Thus, we cannot claim we have determined the exponent in the functional form in Eq.~\eqref{eq:sphal_rate_ans1}, as the explored temperature range is too narrow. We also tried a power-law functional form of the type:
\beq\label{eq:sphal_rate_ans2}
\frac{\Gamma_\S}{T^4} = A \frac{1}{x^C}, \qquad x=T/T_c.
\eeq
Also in this case one finds an excellent description of the data, and an exponent $C\simeq 2.19(38)$. The result is shown in the right panel of Fig.~\ref{fig:sphalrate_vs_T} as a shaded band. Clarifying the actual temperature dependence of the sphaleron rate requires calculations at larger temperatures of the order of $1$ GeV, i.e., $T/T_c \sim \mathcal{O}(10)$.

\section{Conclusions and the road ahead}\label{sec:conclusions}

The results of Ref.~\cite{Bonanno:2023thi} have clearly demonstrated that, using the novel method of~\cite{Bonanno:2023ljc}, one can reliably determine the non-perturbative QCD sphaleron rate from the lattice with an excellent control over several sources of systematic errors (finite lattice spacing, finite smearing width, finite smoothing radius). Now that this new avenue of research has been opened, it would be interesting to further push the first investigation of Ref.~\cite{Bonanno:2023thi} towards the regimes that are interesting for phenomenological applications. Two examples would be to extend the temperature range towards the GeV scale, and to explore the momentum dependence of the topological rate in Eq.~\eqref{eq:topological_rate_nonzeromomenta}, both of which are necessary inputs for axion phenomenology.

To complete these tasks one has to face a serious computational challenge. Indeed, reaching temperatures of the order of 1 GeV on lattices which have at least $N_t\sim 12-16$ temporal sites (a necessary requirement to avoid significant lattice artifacts), requires lattice spacings of the order of $a \sim 0.02 - 0.01$ fm. It is well known that pushing Lattice QCD simulations towards this regime is very hard due to the infamous topological freezing computational problem~\cite{Alles:1996vn,DelDebbio:2002xa,DelDebbio:2004xh,Schaefer:2010hu}. In a few words, when approaching the continuum limit, standard Monte Carlo updating algorithms severely lose ergodicity, especially when focusing on the sampling of the topological charge. In practice, this manifests in a dramatic growth of the number of updating steps necessary to generate two decorrelated samples of the topological charge as $a \to 0$. Apart from this major issue, one should also take into account that the $2+1$ setup employed in Ref.~\cite{Bonanno:2023thi} is not adequate to reach $T \sim 1$ GeV since, at that scale, the dynamical charm quark contribution becomes non-negligible, being $m_{\rm c}/T \sim \mathcal{O}(1)$. The latter point, in particular, implies that one ought to push $2+1+1$ scale setting towards lattice spacings as fine as $a\sim 0.01$ fm to reliably simulate QCD around a 1 GeV temperature, something that is again computationally highly non-trivial (especially because of topological freezing).

\begin{figure}[!t]
\centering
\includegraphics[scale=0.3]{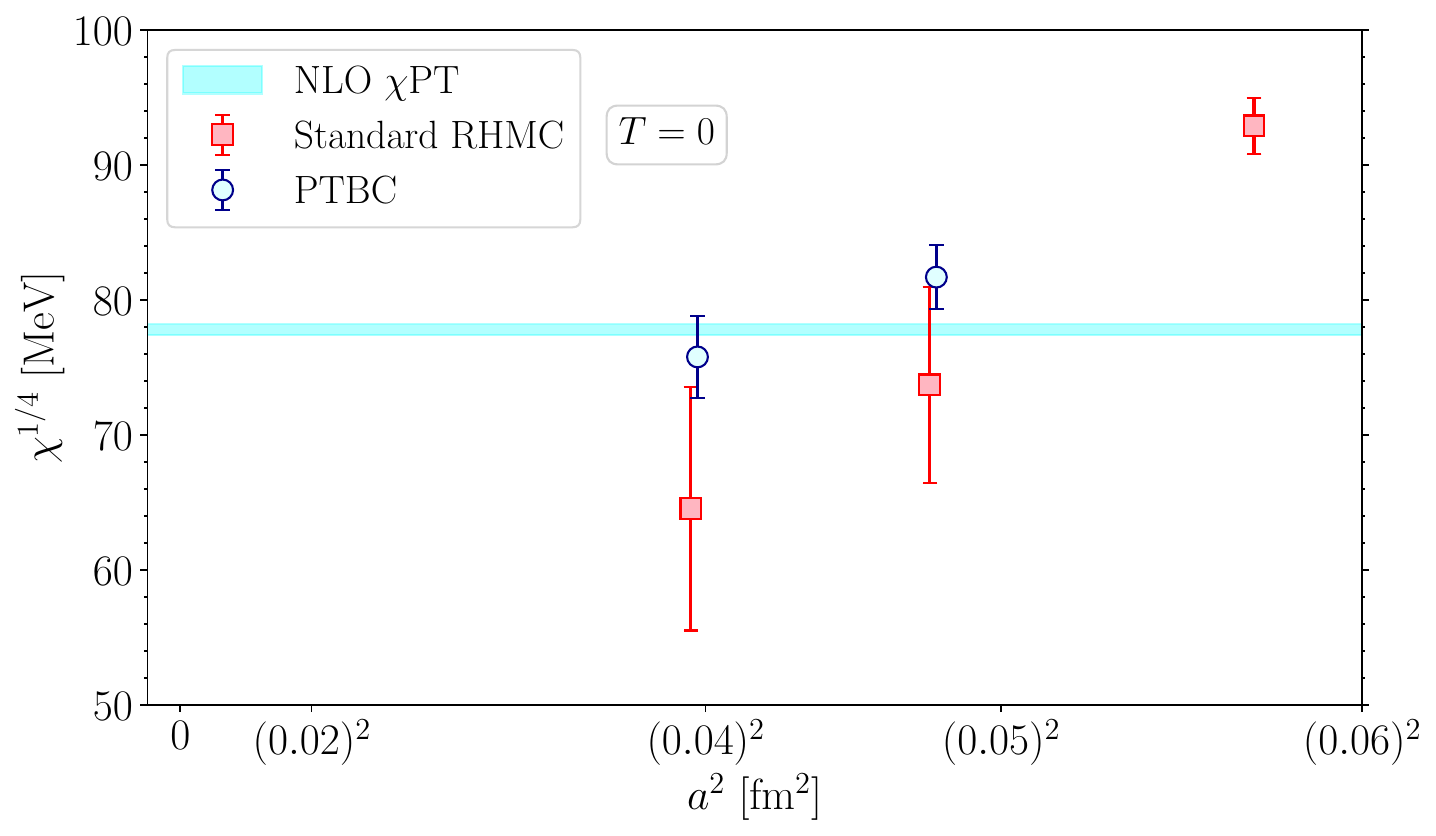}
\includegraphics[scale=0.3]{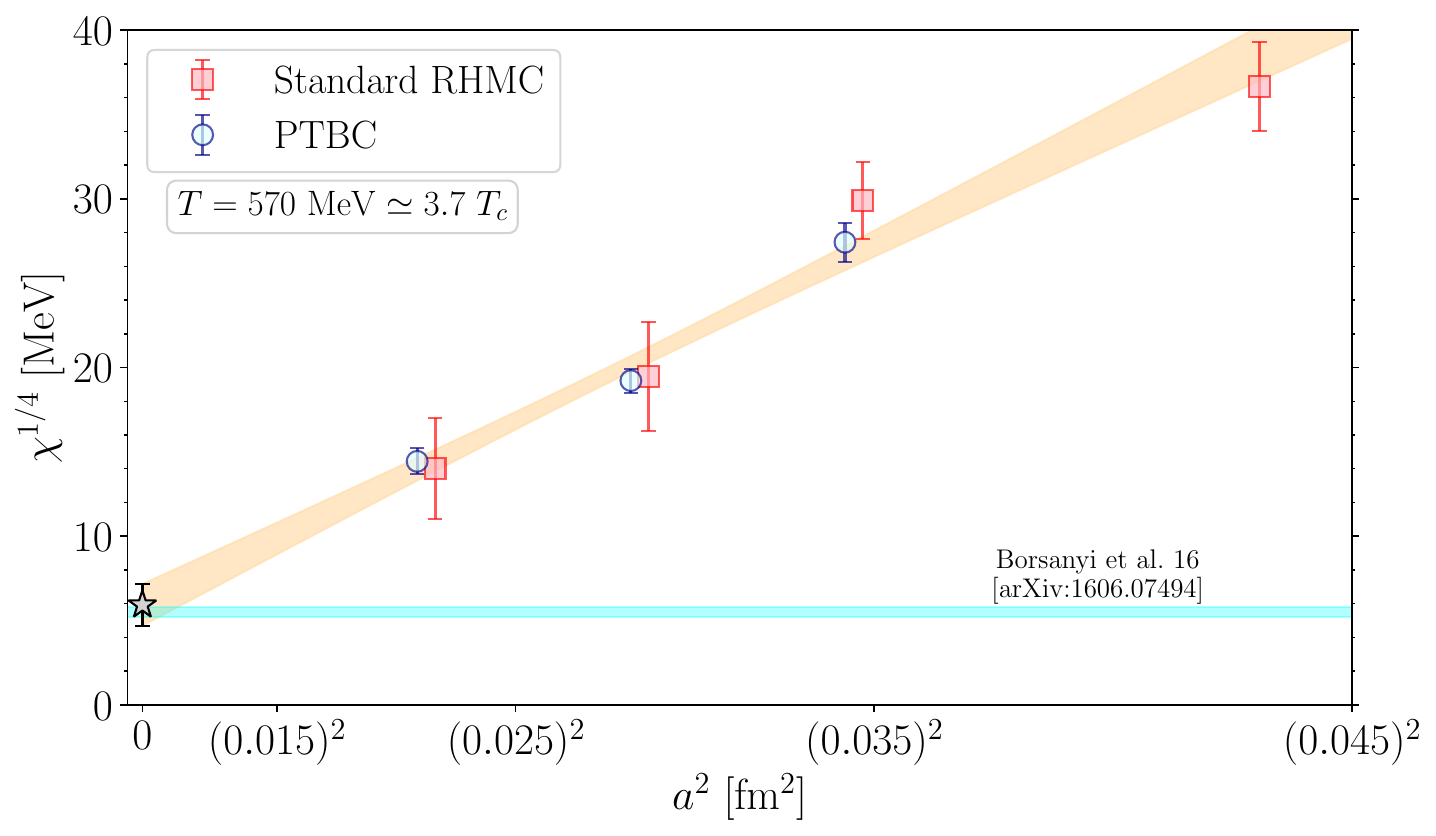}
\caption{Improvement obtained adopting the PTBC algorithm in the determination of the topological susceptibility in $2+1$ QCD at the physical point with respect to the standard RHMC algorithm. Left panel: zero temperature case. Right panel: results for $T=570~\mathrm{MeV}\simeq 3.7 \, T_c$. Figures from Ref.~\cite{Bonanno:2024zyn}.}
\label{fig:topsusc_PTBC}
\end{figure}

In recent years, there were significant efforts in devising new strategies to defeat topological freezing, targeted towards performing efficient simulations on very fine lattices, see, e.g., Refs.~\cite{Giusti:2018cmp,Funcke:2019zna,Kanwar:2020xzo,Albandea:2021lvl,Cossu:2021bgn,Borsanyi:2021gqg,Cranmer:2023xbe,Eichhorn:2023uge,Howarth:2023bwk,Bonanno:2024zyn,Abe:2024fpt,Bonanno:2025pdp}. Among these proposals, me and my collaborators have extensively employed a new algorithmic solution that allows to efficiently avoid topological freezing down to extremely fine lattice spacings: \emph{Parallel Tempering on Boundary Conditions} (PTBC)~\cite{Hasenbusch:2017unr,Bonanno:2020hht, Bonanno:2022yjr, Bonanno:2023hhp, Bonanno:2024nba, Bonanno:2024ggk,Bonanno:2024zyn, Bonanno:2025eeb,Bonanno:2025kfd}. In particular, in the recent study~\cite{Bonanno:2024zyn}, we first implemented it in $2+1$ QCD at the physical point obtaining significant improvements with respect to the standard RHMC. This is clearly illustrated in Fig.~\ref{fig:topsusc_PTBC}, which reports determinations of the QCD topological susceptibility at zero and high temperature via PTBC and standard RHMC.

The left panel of Fig.~\ref{fig:topsusc_PTBC} shows the topological susceptibility in the $T=0$ case, where lattice results can be compared with the prediction obtained from next-to-leading order (NLO) Chiral Perturbation Theory ($\chi$PT)~\cite{GrillidiCortona:2015jxo} (see also~\cite{Gorghetto:2018ocs}). As it can be seen, at the fairly fine lattice spacing $a \simeq 0.056$ fm the lattice result for $\chi$ is still off by about a factor of 2 compared to the NLO $\chi$PT prediction. This is due to the presence of large lattice artifacts affecting $\chi$ in the presence of light dynamical quarks~\cite{Bonati:2015vqz}. Going below this threshold is hard with standard methods due to topological freezing, and this is reflected in the large errors affecting the susceptibility determined for finer lattice spacings with standard methods. The PTBC algorithm instead, significantly reduces autocorrelation times of the topological charge, allowing a reliable determination of $\chi$ in this regime. As it can be seen, PTBC results obtained down to $a \simeq 0.04$ fm are much more accurate, and in nice agreement with the NLO $\chi$PT result. The right panel of Fig.~\ref{fig:topsusc_PTBC} shows instead the finite-temperature case, where the obtained improvement is even more impressive. In that case, we pushed our investigation down to $a\simeq 0.02$ fm, and could achieve a huge improvement in the continuum determination of $\chi$. Our final result is found to be in agreement with the result of~\cite{Borsanyi:2016ksw}, obtained with an indirect method, based on a low-lying Dirac eigenvalue reweighting of the gauge configurations to reduce lattice artifacts by hand.

In conclusion, thanks to the improvements yielded by the PTBC algorithm, I anticipate major progress in the next future about the study of QCD topology from the lattice. These will concern both the study of the topological susceptibility and of the sphaleron rate in the regimes relevant for physical applications, most importantly axion cosmology and collider physics inputs. 

\acknowledgments

\noindent I would like to thank Massimo D'Elia and Margarita Garc\'ia P\'erez for reading this article.\\
I acknowledge support from the Spanish Research Agency (Agencia Estatal de Investigaci\'on) through the grant IFT Centro de Excelencia Severo Ochoa CEX2020-001007-S and, partially, by the grant PID2021-127526NB-I00, both funded by MCIN/AEI/10.13039/501100011033.\\
Numerical simulations of Refs.~\cite{Bonanno:2023ljc,Bonanno:2023thi,Bonanno:2024zyn} have been performed on the \texttt{Marconi}, \texttt{Marconi100}, and $\texttt{Leonardo}$ machines at Cineca, based on the agreement between INFN and Cineca, under projects INF22\_npqcd, INF23\_npqcd, and INF24\_npqcd.

\providecommand{\href}[2]{#2}\begingroup\raggedright\endgroup

\end{document}